\documentclass[graphicx, nofootinbib, notitlepage, twocolumn, amsmath, amssymb, longbibliography]{revtex4-2}
\usepackage{dcolumn}  
\usepackage{bm}       
\usepackage{physics}  
\usepackage{braket}
\usepackage{dsfont}
\usepackage{enumitem}
\usepackage{tikz}
\usepackage{xcolor}
\usepackage{float}

\def\t{\text}

\def\beit{\begin{itemize}}
\def\eit{\end{itemize}}

\begin{document}


\title{Theory of vibronic adsorbate-surface Fano resonances}
\author{Luis Martinez-Gomez}
\author{Sara T. Gebre}
\author{Tianquan Lian}
\author{Raphael F. Ribeiro}
\email[]{raphael.ribeiro@emory.edu}
\affiliation{Department of Chemistry and Cherry Emerson Center for Scientific Computation, Emory University, Atlanta, GA, 30322}

\date{\today}

\begin{abstract}
Inspired by recent visible pump - infrared probe spectra reported for molecular catalysts adsorbed to quantum dots, we introduce a theory of non-equilibrium vibronic Fano resonances arising from the interference of quantum dot excited-state intraband transitions and infrared vibrational excitations of a molecular adsorbate. Our theory suggests a superexchange mechanism for the observed Fano resonances where charge-transfer states mediate the effective interaction between molecular vibrations and near-resonance quantum dot intraband transitions. We present a perturbative treatment of the effective adsorbate-quantum dot vibronic interaction and employ it to construct a two-reservoir Fano model that enables us to capture key experimental trends, including the relationship between the Fano asymmetry factor, quantum dot size, and the distance between the molecular adsorbate and the quantum dot. We focus on adsorbate-quantum dot species, but our theory's implications are shown to be generic for molecules adsorbed to materials with resonant strong mid-infrared electronic transitions.\end{abstract}
\maketitle
\section{Introduction}
\par Fano resonances emerge from the hybridization of interacting discrete and continuum states of a system \cite{fano1961effects} and have been observed in the autoionization spectra of atoms \cite{rice1933predissociation, beutler1935absorptionsserien}, carbon nanotube transport \cite{del2005defective, kim2003fano}, doped-Si Raman scattering \cite{cerdeira1973effect}, scanning tunneling microscopy (STM) measurements of mesoscopic systems with impurities \cite{li1998kondo,madhavan1998tunneling, luo2004fano, ujsaghy2000theory}, and plasmonic materials \cite{luk2010fano, liu2017fano}. Basic interest in Fano resonances stems from their role in fundamental processes underlying energy relaxation and conversion\cite{liu2017electrically, giannini2011fano} and charge transport in materials \cite{miroshnichenko2010fano, xiao2016detecting}. Applications also abound in metamaterials and nanophotonics \cite{song2006tunable, yu2014fano, zhou2014progress, limonov2017fano, gao2018fano, bekele2019plane}, where the high sensitivity of Fano resonances to geometric and environmental changes is promising for a range of applications including sensing \cite{hao2008symmetry, chen2018plasmonic, wu2012fano, king2015fano}, near-field imaging \cite{luk2013fano, song2016nanofocusing, conteduca2023fano}, and nonlinear optics \cite{piao2015spectral, yang2015nonlinear}.

Recently, \citet{yang2021photoinduced} and Gebre et al. \cite{gebre24, gebre24b} reported the observation of Fano-like asymmetric lineshapes in the pump-probe spectrum of inorganic transition metal complexes adsorbed to CdS and CdSe nanocrystals. In these experiments, a visible pump photon generates a QD exciton corresponding to a nanocrystal state with broad infrared adsorption that overlaps with molecular vibrational transitions of the adsorbed molecules. This feature gives rise to the reported Fano resonances observed in the ultrafast infrared probe spectrum, which arise from the interference of the molecular normal-mode transition and the intraband excitation of the photogenerated electron.

In Ref. \cite{gebre24}, the measured Fano asymmetry factors observed for QD-adsorbate complexes show characteristic dependence on the QD size. Similarly, Ref. \cite{gebre24b} reports the variation of the Fano asymmetry with the distance between the QD and the adsorbate. These variations positively correlate with charge-transfer rates and suggest tunable vibronic interactions between the QD and the molecular adsorbate.  In this article, we propose an effective theoretical model and microscopic mechanism for Fano resonances arising from the interference of quantum dot intraband transitions and vibrational excitations of adsorbed molecules. Our theory is validated by comparing its predictions for the size and distance-dependence of experimentally obtained Fano asymmetry factors in recent experiments \cite{gebre24, gebre24b}. In Sec. \ref{sec:theory}, we introduce our theoretical model and describe its underlying assumptions. Sec. \ref{sec:res&dis} provides a derivation of the effective interaction between the molecular vibrations and the QD intraband transitions and the associated vibronic Fano resonance lineshape and discusses its predictions in the context of recent experiments. Sec. \ref{sec:conclusions} summarizes our main results and conclusions. 

\begin{center}
    \begin{figure}[h]
        \includegraphics[width=6.5cm]{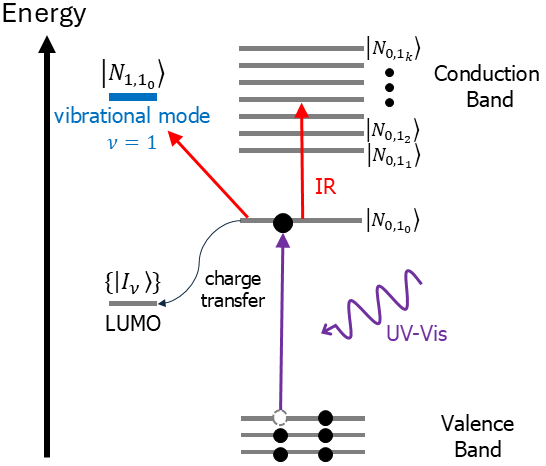}
        \caption{Schematic representation of the vibronic model investigated here where a vibrational excitation of an adsorbed molecule effectively interacts with an intraband QD electronic transition, leading to a Fano resonance. The UV-VIS pulse generates the state $\ket{N_{0,1_0}}$ where the molecular system is in its electronic ground state and the QD is in its lowest energy exciton state.  The IR pulse promotes intraband QD transitions and molecular vibrational excitation, which interfere to give a Fano resonance mediated by virtual charge transfer between the QD and the LUMO.}
        \label{fig:Vibronic}
    \end{figure}
\end{center}

\section{Theory} \label{sec:theory}
We model the quantum dot and its interaction with the molecular system via a tight-binding model. The QD is represented by its 1S and 1P electronic states and bath modes, whereas the molecular electronic Hilbert space is characterized by the occupation of the lowest unoccupied molecular orbital (LUMO). Vibronic coupling in the molecular subsystem is treated using the displaced oscillator model, indicating the ionic molecular species has a different equilibrium geometry than the neutral state. The interaction between the molecular system and the QD is represented by transition matrix elements enabling electron hopping from the QD into the LUMO and vice-versa.  In the following subsections, we describe in detail the molecular, quantum dot, and interaction Hamiltonians employed to derive a vibronic IR Fano lineshape for molecules adsorbed to QDs.

\subsection{Molecular, quantum dot and interaction Hamiltonians}
The total Hilbert space of our system is a direct product of the quantum dot Hilbert space and the molecular vibrational-electronic space of states. The quantum dot electronic states correspond to the lowest energy exciton state, 1S, and the 1P exciton state accessible by IR excitation. The 1S $\rightarrow$ 1P absorption is extremely broad due to the coupling of the 1P exciton with bath degrees of freedom (e.g., quantum dot phonons). The molecular electronic states are characterized by the occupation of the LUMO, and the vibrational degrees of freedom are represented by a single mode assumed to be harmonic for simplicity. Therefore, the quantum states in our system can be classified into neutral ($N$) or ionic ($I$), wherein the neutral manifold, the LUMO state has vanishing occupation number, and the QD is in the 1S or 1P state. The ionic states consist of all levels where an electron occupies the LUMO, and the QD has a single hole in its valence band.

\par We denote the neutral states by $\{ \ket{N_{\nu,n_k}} \}$ corresponding to the quantum dot in electronic state $\ket{n_k}$ ($n_k = 1_k$ if the QD occupies electronic level $k >0$ corresponding to 1P and bath state represented by $k$, and $1_0$ if the QD is in the 1S state), and the molecule is in its electronic ground-state at vibrational level $\nu$.  The initial state (experimentally obtained after pumping the QD using a visible pulse) (Fig. 1) in our analysis is therefore $\ket{N_{0,1_0}}$. The set of ionic states is denoted by $\{ \ket{I_{\nu}}\}$, where $\nu$ labels the molecular vibrational quantum number $\nu$ and the QD is electron deficient. In Fig. \ref{fig:Vibronic}, we provide a schematic representation of our model.

\par Our model has Hamiltonian given by\begin{align}
    H = H_\text{M} + H_{\text{QD}} + H_\text{CT} + H_{\text{ext}}, 
\end{align} where $H_{\text{M}}$ is the molecular part, $H_\text{QD}$ is the QD Hamiltonian, $H_{\text{CT}}$ includes the charge-transfer interaction of the QD with the adsorbate and $H_{\text{ext}}$ is the interaction of the system with an external radiation field. It is useful to partition the Hamiltonian into $H = H_0 + H_{\text{int}} + H_{\text{ext}}$, where $H_0 = H_{\text{M}} + H_{\text{QD}}$ is the noninteracting Hamiltonian describing the isolated QD and molecule,  and $H_{\text{int}}$ contains the charge-transfer interaction and the coupling to the external field.

\par The molecular system is assumed to have distinct equilibrium geometries in the neutral (vanishing LUMO occupation) and ionic states. Therefore, $H_0$ is given by
\begin{align}
    H_0 = &\sum_{\nu} (\epsilon_0+\nu\hbar\omega_M )\ket{N_{\nu,1_0}}\bra{N_{\nu,1_0}} + \nonumber \\
    & \sum_{k\neq 0} \sum_\nu (\epsilon_k+\nu\hbar\omega_M )\ket{N_{\nu,1_k}}\bra{N_{\nu,1_k}} \nonumber \\
    & + \sum_{\tilde{\nu}} E_{I_{\tilde{\nu}}}\ket{I_{\tilde{\nu}}}\bra{I_{\tilde{\nu}}},
\end{align}
where $\nu = 0,1,2,...$ is the vibrational quantum number of the molecular system, $\epsilon_0$ is equal to the sum of the electronic energy of the adsorbate and quantum dot, $\omega_M$ is the molecular vibrational frequency, $\tilde{\nu}$ denotes vibrational quantum number in the ionic state, and $E_{I_{\tilde{\nu}}}$ is the electronic-vibrational energy of the corresponding ionic state. 

\par In what follows, for the sake of simplicity, we will only explicitly consider dynamics involving neutral states $\ket{N_{1,1_0}}$ and $\ket{N_{0,1_k}}$ and their coupling to generic ionic states. These two states are relevant because they correspond to the dominant contributions to the initial and final states of the infrared absorption process of interest. 
\par The interaction Hamiltonian $H_{\text{CT}}$ contains the electronic coupling promoting charge transfer between the QD and molecular species \cite{zhu2016charge, liu2017controlling, marcus1964chemical} and is given by
\begin{align} \label{eq:H_CT}
    H_{\text{CT}} =& \sum_{\tilde{\nu} =0} \left[  t_{I \leftarrow N}^{\tilde{\nu} 0}\op{I_{\tilde{\nu}}}{N_{0,1_0}} + t_{N \leftarrow I}^{0 \tilde{\nu}}\op{N_{0,1_0}}{I_{\tilde{\nu}}} \right.  \nonumber \\
     & \left. +  t_{I \leftarrow N}^{\tilde{\nu} 1} \op{I_{\tilde{\nu}}}{N_{1,1_0}} + t_{N \leftarrow I}^{1 \tilde{\nu}} \op{N_{1,1_0}}{I_{\tilde{\nu}}} \right] \nonumber \\
    \quad & + \sum_{\substack{\tilde{\nu} = 0 \\ k \neq 0}} \left[ t_{I \leftarrow k}^{\tilde{\nu} 0} \op{I_{\tilde{\nu}}}{N_{0,1_k}} + t_{k \leftarrow I}^{0 \tilde{\nu}} \op{N_{0,1_k}}{I_{\tilde{\nu}}} \right],
\end{align}
where $\left(t_{I \leftarrow N}^{\tilde{\nu} 0} \right)^* = t_{N \leftarrow I}^{0 \tilde{\nu}}$ and $\left(t_{I \leftarrow k}^{\tilde{\nu} 0}\right)^* = t_{k \leftarrow I}^{0\tilde{\nu}}$ are vibronic hopping amplitudes between the QD and the molecular LUMO (ionic state) accompanied by a transition into the vibrational state $\tilde{\nu}$ \cite{hestand2018expanded, fulton1961vibronic}, i.e., the electron hopping element $t_{I \leftarrow N}^{\tilde{\nu} \tilde{\nu}'}$ indicates an electron transfer from the neutral to ionic state and a transition of the vibrational molecular states of the form $\tilde{\nu} \leftarrow \tilde{\nu}'$. A detailed description of how the electronic hopping amplitudes are estimated as a function of the various relevant parameters characterizing the QD is given in the next subsection.

\begin{center}
    \begin{figure}
  \includegraphics[width=7cm]{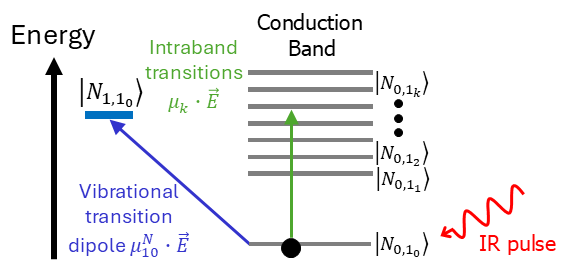}
        \caption{Molecular vibrational (blue arrow), and electronic QD intraband transitions (green arrow) from the lowest state $\ket{N_{0,1_0}}$ induced by the external (infrared) probe electric field.}
        \label{fig:Diagram2}
    \end{figure}
\end{center}

The interaction with the external field models the IR probe pulse induced electronic intraband transition of the QD \cite{h2005infrared} and $0\rightarrow 1$ vibrational transitions of the molecular system in the neutral and ionic states \cite{trenary2000reflection, yang2017ir, persson1981vibrational}
\begin{align}
    H_{\text{ext}} = &- \mu_{10}^N \cdot \vec{E} \op{N_{1,1_0}}{N_{0,1_0}} - \mu_{01}^N \cdot \vec{E} \op{N_{0,1_0}}{N_{1,1_0}} \nonumber \\
    & - \sum_{k\neq 0} \mu_k^{N}\cdot \vec{E}\left(\ket{N_{0,1_k}}\bra{N_{0,1_0}} \ket{N_{0,1_0}}\bra{N_{0,1_k}} \right) \nonumber \\
    & - \sum_{ \substack{\nu = 0 \\ \nu' > \nu}} \left( \mu_{\nu \nu'}^I \cdot \vec{E} \op{I_\nu}{I_{\nu'}} + \mu_{\nu' \nu}^I \cdot \vec{E} \op{I_{\nu'}}{I_\nu} \right) 
\end{align}
Here, $\vec{E}$ represents the IR electric field, and $\mu_{10}^N$ ($\mu_{\mu'\mu}^I$) denotes the vibrational transition dipole matrix element between the vibrational states $0 \rightarrow 1$ ($\mu \rightarrow \mu'$) when the molecular-QD system is in the neutral (ionic) state, whereas $\mu_k^{N}$ is the QD electronic intraband transition dipole moment for the lowest energy exciton state into the $k$ state. In Figure \eqref{fig:Diagram2}, we show a schematic representation of the electronic and vibrational transitions induced by the external field (IR probe pulse).

\subsection{Electronic coupling} \label{subsec:EC}
\par The electronic coupling between the QD surface and the adsorbed molecule, denoted by $t_{A \leftarrow B}$, is determined by the electronic interaction between both species
\begin{align}
    t_{I \leftarrow N} = \int_{-\infty}^{\infty}  \psi^*_{\text{M}}(\vec{r})H_{\text{int}} \Psi_{\text{QD}}(\vec{r}) \dd \vec{r},
\end{align}
where $\psi_M$  is the electronic wavefunction of the LUMO, $\Psi_{\text{QD}}$ is a QD state, and $H_{\text{int}}$ is the (mean-field) electronic interaction Hamiltonian. For the sake of simplicity, since our analysis below is focused on investigating variations of the QD-adsorbate effective interaction with QD properties independent of molecular vibrational degrees of freedom, we ignore the vibrational part of the adsorbate and QD wave function in this section. 

\par The penetration amplitude of the molecular wavefunction into the QD bulk is negligible and, therefore, neglected. In this case, the electronic overlap is only significant near the surface \cite{zeiri2019third}, and the electronic hopping integral can be written as
\begin{align}
    t_{I \leftarrow N} = \int_{R_0}^{\infty} \psi_{\text{M}}^*(\vec{r}) H_{\text{int}} \Psi_{\text{QD}}(\vec{r}) \dd \vec{r},
\end{align}
where $R_0$ is the QD radius. Assuming a perfectly spherical QD and using Eq. \eqref{eq:el_wave}, we find
\begin{align} \label{eq:t_hopping}
    t_{I \leftarrow N} & = \int_{R_0}^{\infty} \psi_{\text{M}}^*(r)H_{\text{int}} R_{nl}(r) r \dd{r}.
    \end{align}
This implies the electronic hopping parameter $t_{I \leftarrow N}$  is approximately proportional to  $R_0R_{nl}(R_0)$. Given that $R_{nl}(R_0)$ decays exponentially and thus much faster than $R_0$ increases with the QD size, the charge transfer amplitude $t_{I \leftarrow N}$ is anticipated to become less efficient with increasing QD size.

\begin{center}
    \begin{figure}[h]
   \includegraphics[width=8cm]{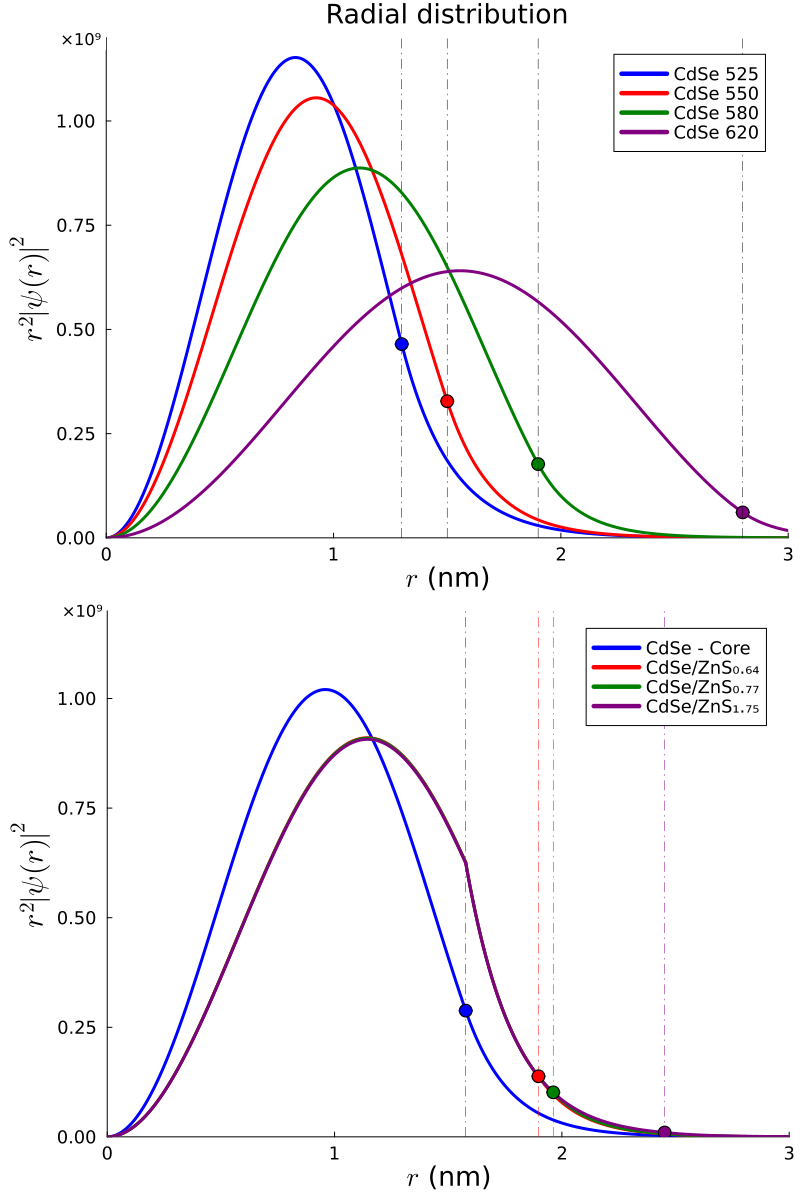}
        \caption{Radial distribution of a 1S wavefunction for QDs with sizes of interest to our work and investigated experimentally in Refs. \cite{gebre24} and \cite{gebre24b}. In (a), we show the radial distribution at various CdSe sizes, whereas (b) shows the 1S radial distribution for different CdSe/ZnS Core/Shell thicknesses with fixed core sizes.}
        \label{fig:wavefunction}
    \end{figure}
\end{center}

In the experiments discussed in Ref. \cite{gebre24b}, core/shell QDs were employed to modulate the interaction between the QD and the molecular adsorbate. Specifically, the shell thickness enabled control of the overlap between the adsorbate and the QD exciton wave functions. Appendix A describes our procedure for estimating the exciton wave function in core/shell QD structures.
In Fig. \ref{fig:wavefunction}, we show the obtained radial distribution of 1S electronic wave functions of the core QD at various radii and for core/shell structures at different shell thicknesses. 

\section{Results and Discussion} 
In this section, we employ the Hamiltonian described in the previous section and perturbation theory to describe the emergence of an effective interaction between QD intraband transitions and adsorbate vibrations via virtual charge transfer processes. This interaction is then applied in a two-reservoir Fano model to characterize the behavior of previously reported variation with QD size and molecular distance of IR Fano lineshapes.
\label{sec:res&dis}
\subsection{Vibronic Fano resonances}
In our vibronic model, the relevant part of the Hilbert space for analysis of the IR adsorbate-QD Fano resonances in Refs. \cite{gebre24} and \cite{gebre24b} is spanned by the 0th-order molecular vibrational state $\ket{N_{1,1_0}}$ (the discrete final state), the electronic states of the quantum dot (QD)  embedded in the continuum $\ket{N_{0,1_k}}$, the lowest energy electronic state in the vibrational ground-state  $\ket{N_{0,1_0}}$, and the set of ionic states $\ket{I_{\tilde{\nu}}}$ (ReC0A$^-$ QD$^+$) with variable vibrational quantum number $\nu$. The ionic states weakly mix with the neutral via the charge-transfer interaction $H_{\text{CT}}$ and inspired by first-order perturbation theory, we define a new basis for neutral and ionic species where the relevant states in what follows are given by
\begin{align} 
    \ket{\tilde{N}_{0,1_0}} = & A_{0,1_0}^N \left( \ket{N_{0,1_0}} + \sum_{\tilde{\nu} = 0} \frac{t_{I \leftarrow N}^{\tilde{\nu} 0}}{\epsilon_0 - E_{I_{\tilde{\nu}}}}\ket{I_{\tilde{\nu}}}\right), \label{eq:N010} \\
    \ket{\tilde{N}_{1,1_0}} = & A_{1,1_0}^N \left( \ket{N_{1,1_0}}+ \sum_{\tilde{\nu} = 0} \frac{t_{I \leftarrow N}^{\tilde{\nu} 1}}{\hbar \omega + \epsilon_0 - E_{I_{\tilde{\nu}}}}\ket{I_{\tilde{\nu}}} \right), \label{eq:N110} \\
    \ket{\tilde{N}_{0,1_k}} = & A_{0,1_k}^N \left( \ket{N_{0,1_k}} + \sum_{{\tilde{\nu}} = 0} \frac{t_{I \leftarrow k}^{{\tilde{\nu}} 0}}{\epsilon_k - E_{I_{\tilde{\nu}}}}\ket{I_{\tilde{\nu}}} \right), \label{eq:N01k} 
\end{align}
and the $A$ parameters are normalization coefficients. Using these new states (and their orthogonal complement), we obtain a unitarily transformed Hamiltonian $\tilde{H}$ with  effective interactions $\tilde{C}_k$ between the near-resonance vibrational excitation $\ket{\tilde{N}_{1,1_0}}$ and the excited QD states $\ket{\tilde{N}_{0,1_k}}$ (see Appendix B for further details) suggesting hybridization between discrete and continuum levels with eigenstates 
\begin{align}
    \ket{\tilde{\Psi}} = a_0 \ket{\tilde{N}_{1,1_0}} + \sum_k b_k \ket{\tilde{N}_{0,1_k}}.
\end{align}
connected to the initial-state $\ket{\tilde{N}_{0,1_0}}$ via the external radiation field and $H_{\t{ext}}$. It follows the probability of excitation between the initial state $\ket{\tilde{N}_{0,1_0}}$ and the stationary $\ket{\tilde{\Psi}}$ leads to the Fano lineshape
\begin{align}\label{eq:Fano_Lineshape}
    \frac{ \abs{ \mel*{ \tilde{ \Psi } }{ H_{ \text{ext} } }{ \tilde{N}_{0,1_0} } }^2}{ \abs{ \mel*{ \tilde{N}_{0,1_k} }{ H_{ \text{ext} } }{ \tilde{N}_{0,1_0} } }^2} = \frac{\abs{q + \epsilon}^2}{\epsilon^2 + \xi + 1},
\end{align} 
where $\epsilon$ is the reduced IR photon energy, $q$ is the Fano asymmetry parameter, and $\xi$ denotes the number of levels of the discretized continuum within the natural width. The Fano asymmetry parameter is given by"
\begin{align} \label{eq:Fano_q}
    q &= \frac{ \tilde{W}_v }{ \tilde{W}_e \tilde{C}_k + W_{\bar{\omega}}V_{\bar{\omega}}  } \cdot \frac{ \abs{ \tilde{C}_k }^2 + \abs{V_{\bar{\omega}}}^2 }{ (\hbar \Gamma/2) },\\
    & = \frac{1}{\pi\rho(\omega)} \cdot \frac{ \tilde{W}_v }{ \tilde{W}_e \tilde{C}_k + W_{\bar{\omega}}V_{\bar{\omega}}} \label{eq:Fano_q1}
\end{align}
where $\tilde{W}_v$ is the vibrational transition dipole moment matrix element, $\tilde{C}_k$ is the effective coupling between the molecular vibrations and the continuum QD electronic states, and $\tilde{W}_e$ is the 1S $\rightarrow$ 1P electronic transition dipole moment. The parameters $V_\omega$  and $W_\omega$ model the interaction of the relevant adsorbate vibrational mode with other degrees of freedom that lead to homogeneous broadening of the molecular vibrational excitation and its characteristic Lorenzian lineshape which is present even when the QD is in its ground-state (prior to UV excitation in the experiments of Ref. \cite{gebre24} and \cite{gebre24b}). In the presence of the effective coupling to the QD $1S \rightarrow 1P$ transition, the vibrational dephasing rate is therefore estimated by
\begin{align} \label{eq:Gamma_FGR}
    \Gamma = \frac{2 \pi}{\hbar} \left( \abs{\tilde{C}_k}^2 + \abs{V_{\bar{\omega}}}^2 \right) \rho(\omega),
\end{align}
where $\rho(\omega)$ is the average density of states of the two reservoirs. The effective coupling of the second reservoir is expected to be greater since the experiments of Refs. \cite{gebre24} and \cite{gebre24b} show the molecular vibrational lineshape is barely affected by the coupling to the excited QD, and in fact, the dominant effect of the latter is in renormalizing the infrared oscillator strength of the vibrational transition. Thus, the effective coupling with the intramolecular or solvent degrees of freedom $W_{\bar{\omega}}V_{\bar{\omega}}$ is expected to be much greater than the QD electronic intraband transitions and $W_{\bar{\omega}}V_{\bar{\omega}} > \tilde{W}_e \tilde{C}_k$.  Figure \ref{fig:Diagram4} shows a schematic representation of the Fano parameters and the states involved in the vibronic Fano resonance.  

\begin{center}
    \begin{figure}[h]
   \includegraphics[width=7cm]{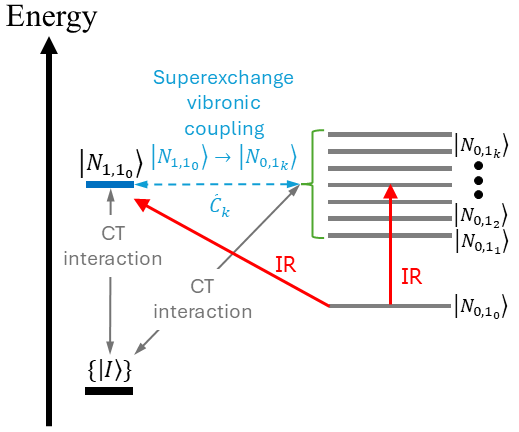}
        \caption{Superexchange vibronic Fano resonance between molecular vibration $\ket{N_{1,1_0}}$ and QD intraband electronic states $\ket{N_{0,1_k}}$ via virtual charge-transfer (CT) interaction of LUMO-QD}
        \label{fig:Diagram3}
    \end{figure}
\end{center}

We present in Appendix \ref{appendix:QFR} detailed explicit derivations of all parameters in Eq. \eqref{eq:Fano_q}. Here, we highlight the effective interaction of the discrete state (molecular vibrations) with the electronic QD continuum states (IR intraband transitions) and denoted by $\tilde{C}_k$ are due to a superexchange process mediated by the CT interaction
\begin{align} \label{eq:Ck_norm}
    \tilde{C}_k \approx 2 \sum_{\tilde{\nu}=0}\frac{t_{k\leftarrow I}^{0 \tilde{\nu}}t_{I \leftarrow N}^{\tilde{\nu} 1}}{\hbar \omega + \epsilon_0 - E_{I_{\tilde{\nu}}}}.
\end{align}
This interaction is schematically illustrated in Fig. \ref{fig:Diagram3}.

We utilize the expressions in Eqs. \eqref{eq:We_norm}-\eqref{eq:Wv_norm} and Eq. \eqref{eq:Ck_norm} to derive the Fano asymmetry lineshapes described by Equation \eqref{eq:Fano_q}.  To simplify the Fano parameters, we employ the following approximations highlighting their dependence on the electronic coupling strength, denoted by $t$.
    
First, we address the light-induced interaction of the initial state $\ket{\tilde{N}_{0,1_0}}$ with the molecular vibration $\ket{\tilde{N}_{1,1_0}}$ as 
\begin{align}
    \tilde{W}_v & \approx \bar{\mu}_{10}^N \cdot \vec{E} + 2\eta_\nu^I t^2 \left(\bar{\mu}_{\nu}^I\cdot \vec{E} \right)  \label{eq:simpl2},
\end{align}
where $\eta_\nu^I t^2 \left(\bar{\mu}_{\nu}^I\cdot \vec{E} \right)$ captures the contributions of the vibrational transition dipole moment induced by the weak admixture with the ionic state:
\begin{align}
    \eta_\nu^I t^2 \left(\bar{\mu}_{\nu}^I\cdot \vec{E} \right) =& \sum_{ \substack{{\tilde{\nu}} = 0 \\ {\tilde{\nu}}' > {\tilde{\nu}}} }\left[ \frac{t_{N \leftarrow I}^{1{\tilde{\nu}}}\mu_{{\tilde{\nu}} {\tilde{\nu}}'}^I t_{I \leftarrow N}^{{\tilde{\nu}}'0}}{(\hbar \omega + \epsilon_0 - E_{I_{\tilde{\nu}}})(\epsilon_0 - E_{I_{{\tilde{\nu}}'}})} \right.  \nonumber \\
    & \left. + \frac{t_{N \leftarrow I}^{1{\tilde{\nu}}'}\mu_{{\tilde{\nu}}' {\tilde{\nu}}}^I t_{I \leftarrow N}^{{\tilde{\nu}}0}}{(\hbar \omega + \epsilon_0 - E_{I_{{\tilde{\nu}}'}})(\epsilon_0 - E_{I_{{\tilde{\nu}}}})} \right] \cdot \vec{E}.
\end{align}
Second, in the regime of small electronic hopping, the intraband transitions involving the lowest QD exciton state and the continuum states ($\tilde{W}_e$) are primarily governed by $\mu_{k}^N$. Consequently, we assume $\tilde{W}_e$ is independent of the CT process, simplifying it to $W_e$ (interaction from the initial state to the QD 1P states in the unperturbed basis).

Lastly, we approximate the summation over all intermediate ionic states contributing to $\tilde{C}_k$ by a constant. This simplification highlights the dominant dependence of $\tilde{C}_k$ on the electronic coupling strength, $t^2$. Consequently, in the Fano asymmetry factor $q$ in Eq. \eqref{eq:Fano_q1}, we replace   $\tilde{W}_e \tilde{C}_k$ by $\bar{W}_{e,C}t^2$ with $\bar{W}_{e,C}$ a dimensionless quantity.
\par Substituting these approximations into the Fano asymmetry factor $q$  given by Eq. \eqref{eq:Fano_q1} we obtain
\begin{align} \label{eq:aprox_q}
    q & \approx \frac{1}{\pi \rho(\omega_0)} \frac{\bar{\mu}_{10}^N  +\eta_\nu^I t^2 \bar{\mu}_{\nu}^I }{\bar{W}_{e,C} t^2 + W_M}, \end{align}
 where we introduced the notation $\bar{\mu}_{10}^N \equiv \bar{\mu}_{10}^N \cdot \vec{E}$ ($\bar{\mu}_{\nu}^I \equiv \bar{\mu}_{\nu}^I \cdot \vec{E}$). In Eq. \eqref{eq:aprox_q}, $W_{M} = W_{\bar{\omega}}V_{\bar{\omega}}$ quantifies the vibrational dephasing into the molecular reservoir present even when the QD is in its ground-state, and here we have also assumed that the average density of states $\rho(\omega_0)$ is independent of energy and remains constant across various QD sizes. This assumption is consistent with our approximation that the continuum density of states represents the average of the molecular and QD and is independent of the QD size or the molecular distance to the QD.

 \par Eq. \ref{eq:aprox_q} is one of the main results of this work.  It expresses the Fano asymmetry factor $q$ in terms of the molecular-QD electron hopping amplitude $t$. This quantity plays a key role in setting the Fano asymmetry factor because it (1) renormalizes the vibrational oscillator strength, substantially enhancing the infrared absorption strength relative to the unperturbed neutral state, and (2) it also controls how much vibrational energy is dumped into the QD in comparison to the molecular reservoir. Below, we discuss the implications of our formalism.

\subsection{Size and distance dependence of Fano resonance}
In this subsection, we connect our theoretical findings on the Fano asymmetry factor $q$ (Eq. \ref{eq:aprox_q}) with experimental data by exploring its dependence on the QD size and shell thickness. The molecular parameters $\bar{\mu}_{\t{Ionic}}(t_{\t{max}})/\bar{\mu}_{10}^N $ and $W_M/W_{\t{QD}}(t_{\t{max}})$ were assumed independent of system size and adjusted by fitting them to experimental data using a distance minimization algorithm. Our analysis focuses on how the electronic coupling between the LUMO of the molecular adsorbate and the QD influences the Fano resonance, with particular attention to the contributions of vibrational transition dipoles and decay rates for CdSe and CdSe/ZnS quantum dots.

\begin{center}
    \begin{figure}[t]
        \includegraphics[width=7.5cm]{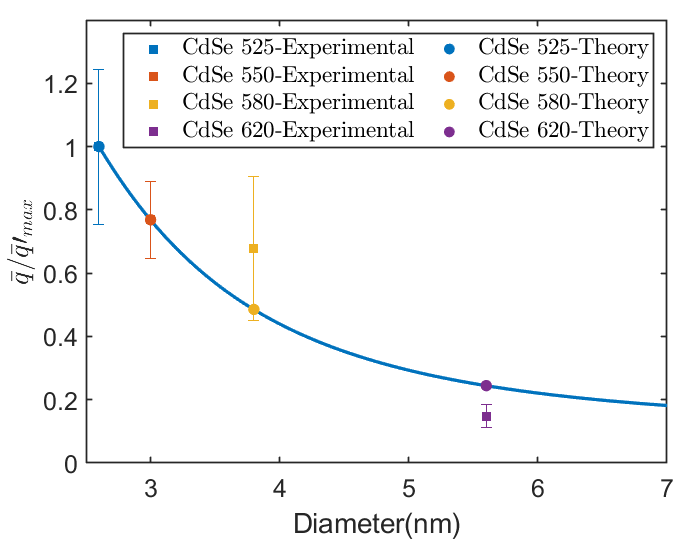}
        \caption{Normalized Fano resonance asymmetry parameter dependence on QD size with $\bar{\mu}_{\t{Ionic}}(t_{\t{max}})/\bar{\mu}_{10}^N = 10$ and $W_M/W_{\t{QD}}(t_{\t{max}}) = 5.21$.  Electronic couplings were normalized relative to the case with the largest $q$ for which we took $t_{\t{max}} = 1$.}
        \label{fig:size_dep_q}
    \end{figure}
\end{center}

Figure \ref{fig:size_dep_q}a reveals agreement between our theoretical predictions and experimental observations regarding the variation of $q$ with QD size, where the $q$ factor decreases as the QD size increases. The size-dependent behavior of $q$ is primarily governed by the electronic coupling $t$ between the exciton state and the molecular LUMO, which is modulated by the QD size. A closer to symmetric absorption ($q \rightarrow \infty$) indicates an enhanced electronic coupling between the QD exciton and ionic electronic molecular states. Notably, the vibrational transition dipole in the ionic state of the molecular-QD system plays a dominant role in determining $q$ (second term in Eq. \ref{eq:simpl2}), highlighting the complex interplay between electronic and vibrational dynamics in these nanoscale systems. Finally, the molecular dephasing, in principle, induced by interaction with the QD electronic transition and other degrees of freedom, is predominantly governed by the molecular mechanism consistent with the approximations in Eq. \ref{eq:Gamma_FGR}.

\par
\begin{figure}[h]
 \includegraphics[width=8.2cm]{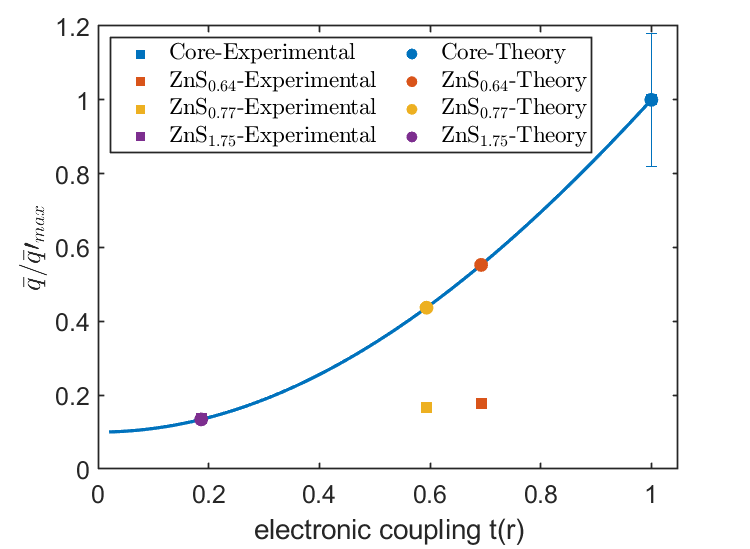}
    \caption{Normalized IR Fano resonance asymmetry parameter variation with the thickness of a core/shell QD. Parameters included $\bar{\mu}_{\t{Ionic}}(t_{\t{max}})/\bar{\mu}_{10}^N = 9.6$ and $W_M/W_{\t{QD}}(t_{\t{max}}) = 10$. Electronic couplings $t$ were normalized relative to the case with largest $q$ for which we took $t_{\t{max}} = 1$.}
    \label{fig:dis_dep_q}
\end{figure}

 In Fig. \ref{fig:dis_dep_q}, we show the behavior of the Fano asymmetry parameter in core/shell QDs with variable thickness. The figure indicates a clear decrease in the normalized Fano asymmetry factor with increasing shell thickness (0, 1, 3, and 6 monolayers) in CdSe/ZnS core/shell QDs, in reasonable agreement with experimental findings. The observed error margin between experimental and theoretical Fano asymmetry parameter dependence on adsorbate-QD distance is largely due to the challenges in modeling the 1S excitonic wavefunction in core/shell QDs and may also be due to potential variation of bare ionic and neutral molecular vibrational transition dipole moments with the core/shell structure. Despite these approximations, the observed trend indicates a weakening of the QD-adsorbed molecule electronic coupling as the shell thickness increases (Fig. \ref{fig:dis_dep_q}b) in agreement with the behavior observed with charge-transfer rates \cite{gebre24b}.

\section{Conclusions} \label{sec:conclusions}
This work introduced a superexchange mechanism for vibronic Fano resonances observed in visible pump-IR probe measurements of molecules adsorbed to QDs. Our proposed mechanism involves coupling the molecular CO vibration with the electronic conduction states of the QD through virtual charge-transfer interactions mediated by vibronic coupling with the LUMO. By incorporating the vibronic coupling of the exciton state with the LUMO of the molecular adsorbate, we obtained an approximate expression for the Fano asymmetry parameter $q$ in terms of the charge-transfer coupling $t$ (Eq. \eqref{eq:aprox_q}). This model successfully captures several experimental observations across various QD configurations, including the reported decrease in the Fano asymmetry $q$ factor with increasing QD size and distance to the adsorbate—reflecting the reduced electronic coupling between the QD exciton and the vibrational molecular states. 

\par The microscopic mechanism provided for vibronic Fano resonances in this work applies broadly and is expected to be observable in visible pump-IR probe measurements whenever molecular vibrations are near-resonant with infrared electronic transitions of QDs or other nanoparticles.  Future experiments and ab initio quantum chemistry simulations provide promising avenues for unraveling further details of the Fano resonances discussed in this work, including the role of the solvent interactions with the QD and the molecular adsorbate.\\

\textbf{Acknowledgments.} This material is based upon work supported by the U.S. Department of Energy, Office of Science, Office of Basic Energy Sciences, Solar Photochemistry Program under Award Number DE-SC0008798. R.F.R acknowledges support from NSF CAREER award Grant No. CHE-2340746 and startup funds provided by Emory University. S.T.G acknowledges support from AGEP supplement to NSF award number CHE-2004080. 

\appendix
\section{Quantum dot wave functions}
The Schrodinger equation for the electron-hole quantum dot wave function $\Psi(r,\theta,\phi)$ and its corresponding energies $E$ can be written in spherical coordinates as:
\begin{align} \label{eq:Schrodinger}
    \left\{ \frac{\hbar^2}{2m_i^*r^2}\left[ \pdv{r} \left( r^2 \pdv{r} \right) + \frac{1}{\sin \theta}\pdv{\theta}\left( \sin \theta \pdv{\theta} \right) \right. \right. \nonumber \\
    \left. \left. + \frac{1}{\sin^2 \theta}\pdv[2]{\phi} \right] + V(r) \right\} \Psi(r, \theta,\phi) = E \Psi(r,\theta,\phi),
\end{align}
where $m_i^*$ is the electron effective mass in the conduction band in the $i$-th layer on the QD, and we have assumed a radial dependence on the potential energy $V(r)$. Note that the conduction band edge energy determines the potential energy \cite{jasieniak2011size}. Thus, we define the potential energy for the bare QD with radius $R_0$ as:
\begin{align} \label{eq:V(r)_bare}
    V(r) = \left\{ \mqty{V_0 & 0<r<R_0\\ 0 & r>R_0} \right.
\end{align}
with effective mass  $m_i^* = m_1^*$. In the Core/Shell structure with inner radius $R_0$ and total radius $R_1$, we define the potential energy as
\begin{align}\label{eq:V(r)_CS}
    V(r) = \left\{ \mqty{V_0 & 0<r<R_0\\ V_1 & R_0<r<R_1\\ 0 & r>R_1,} \right.
\end{align}
and effective mass
\begin{align}
    m_i^* = \left\{ \mqty{m_1^* & 0<r<R_0\\ m_2^* & R_0<r<R_1\\ m_1^* & r>R_1.} \right.
\end{align}
Since we have assumed a perfect spherical QD, and given the potentials \eqref{eq:V(r)_bare}-\eqref{eq:V(r)_CS}, we can split the wavefunction into its radial $R_{nl}(r)$ and angular (spherical harmonics function) $Y_{lm}(\theta,\phi)$ parts:
\begin{align} \label{eq:el_wave}
    \Psi(r,\theta,\phi) = R_{nl}(r)Y_{lm}(\theta,\phi),
\end{align}
where $\theta$ is the polar and $\phi$ is the azimuthal angle. The detailed solutions of Eq. \eqref{eq:Schrodinger} can be found in \cite{zeiri2019third} with the following boundary conditions in the interface between the Core/Shell:
\begin{align}
    R_{nl,i}(R_i) &= R_{nl,i+1}(R_i),\\
    \frac{1}{m_{i}^*} \pdv{r} \eval{R_{nl,i}(r)}_{r=R_i} &= \frac{1}{m_{i+1}^*}\pdv{r}\eval{R_{nl,i+1}(r)}_{r=R_i}.
\end{align}
The QD size and solvent have been shown to influence the edge band energy. For instance, \citep{jasieniak2011size} proposed an empirical formula to calculate the conduction band edge energy for CdSe QD given by:
\begin{align}
    V_0 =  -3.49 + 2.97D^{-1.24},
\end{align}
$D$ is the QD diameter, and $-3.49$ is the bulk conduction band edge energy relative to a solvent with dielectric constant $2$. The conduction band edge position of the complex ZnS relative to the CdSe in a core/shell QD ranges from $V_0+1.4$ to $V_0+0.9$. The values reported here ($V_1 = V_0+1.4$) are chosen to emphasize the shell thickness dependence on the Fano asymmetry factor.\\

\section{Quantum Fano resonance parameters from perturbation theory} \label{appendix:QFR}
\par The Fano parameters of \eqref{eq:Fano_q} are described by the set of perturbed eigenstates \eqref{eq:N010}-\eqref{eq:N01k} interacting via vibronic coupling with the LUMO. Thus, the Fano parameters are defined as follows:
\begin{align}
    \tilde{W}_e & = \bra{\tilde{N}_{0,1_k}}H_{\text{LM}}\ket{\tilde{N}_{0,1_0}}, \label{eq:We_p} \\
    \tilde{W}_v & = \bra{\tilde{N}_{1,1_0}}H_{\text{LM}}\ket{\tilde{N}_{0,1_0}}, \label{eq:Wv_p}\\
    \tilde{C}_k & = \bra{\tilde{N}_{0,1_k}}H_{\text{CT}}\ket{\tilde{N}_{1,1_0}}  \label{eq:Ck_p}.
\end{align}
The Fano parameters \eqref{eq:We_p}-\eqref{eq:Ck_p} can be written by replacing the perturbed eigenstates \eqref{eq:We_norm}-\eqref{eq:Ck_norm} and ignoring higher orders of $O(t^2)$ on the electron hopping as: 
\begin{widetext}
\begin{align}
    \tilde{W}_e \approx & \mu_k^N \cdot \vec{E} \Bqty{ 1 - \frac{1}{2} \sum_{{\tilde{\nu}} = 0} \left[ \abs{ \frac{t_{I \leftarrow k}^{{\tilde{\nu}} 0}}{\epsilon_k - E_{I_{\tilde{\nu}}}} }^2 + \abs{ \frac{t_{I \leftarrow N}^{{\tilde{\nu}} 0}}{\epsilon_0 - E_{I_{\tilde{\nu}}}} }^2 \right] } \nonumber \\ 
    & + \sum_{ \substack{{\tilde{\nu}} = 0 \\ {\tilde{\nu}}' > {\tilde{\nu}}} } \left[\frac{t_{k \leftarrow I}^{0{\tilde{\nu}}} \mu_{{\tilde{\nu}} {\tilde{\nu}}'}^{I} t_{I \leftarrow N}^{{\tilde{\nu}}'0}}{(\epsilon_k - E_{I_{\tilde{\nu}}})(\epsilon_0 - E_{I_{{\tilde{\nu}}'}})} + \frac{t_{k \leftarrow I}^{0{\tilde{\nu}}'} \mu_{{\tilde{\nu}}' {\tilde{\nu}}}^{I} t_{I \leftarrow N}^{{\tilde{\nu}}0}}{(\epsilon_k - E_{I_{\tilde{\nu}}'})(\epsilon_0 - E_{I_{{\tilde{\nu}}}})} \right] \cdot \vec{E} \label{eq:We_norm}, \\
    \tilde{W}_v \approx & \mu_{10}^N \cdot \vec{E} \Bqty{ 1 - \frac{1}{2} \sum_{{\tilde{\nu}} = 0} \left[ \abs{ \frac{t_{I \leftarrow N}^{{\tilde{\nu}} 1}}{\hbar \omega + \epsilon_0 - E_{I_{\tilde{\nu}}}} }^2 + \abs{ \frac{t_{I \leftarrow N}^{{\tilde{\nu}} 0}}{\epsilon_0 - E_{I_{\tilde{\nu}}}} }^2 \right]  }  \nonumber \\
    & + \sum_{ \substack{{\tilde{\nu}} = 0 \\ {\tilde{\nu}}' > {\tilde{\nu}}} }\left[ \frac{t_{N \leftarrow I}^{1{\tilde{\nu}}}\mu_{{\tilde{\nu}} {\tilde{\nu}}'}^I t_{I \leftarrow N}^{{\tilde{\nu}}'0}}{(\hbar \omega + \epsilon_0 - E_{I_{\tilde{\nu}}})(\epsilon_0 - E_{I_{{\tilde{\nu}}'}})} + \frac{t_{N \leftarrow I}^{1{\tilde{\nu}}'}\mu_{{\tilde{\nu}}' {\tilde{\nu}}}^I t_{I \leftarrow N}^{{\tilde{\nu}}0}}{(\hbar \omega + \epsilon_0 - E_{I_{{\tilde{\nu}}'}})(\epsilon_0 - E_{I_{{\tilde{\nu}}}})} \right] \cdot \vec{E} \label{eq:Wv_norm},\\
\end{align}
\end{widetext}

\begin{center}
    \begin{figure}
        \includegraphics[width=7cm]{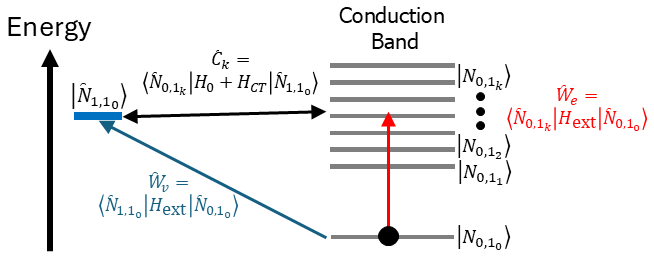}
        \caption{Vibronic Fano mode. Schematic of the Fano parameters in Eq. \eqref{eq:Fano_q}.}
        \label{fig:Diagram4}
    \end{figure}
\end{center}

Equations \eqref{eq:Wv_norm}-\eqref{eq:Ck_norm} delineate the Fano lineshape parameter, highlighting its dependence on the electronic hopping terms $t_{I \leftarrow N}^{{\tilde{\nu}} {\tilde{\nu}}'}$ and $t_{I \leftarrow k}^{{\tilde{\nu}} {\tilde{\nu}}'}$. Fig. \ref{fig:Diagram4} illustrates the Fano parameters and their states involved in the interaction. Note that the Fano parameters described in Eq. \eqref{eq:Fano_q} corresponds to the perturbed eigenstates induced by vibronic coupling.

In Eq. \eqref{eq:Ck_norm} we have established the coupling between molecular vibrations and the discrete states of the QD, denoted by $C_k$, through the electronic coupling mediated by the virtual molecular ionic state (LUMO). This interaction occurs through the following: 1) the molecular vibrational modes in the neutral species interact with the virtual Ionic states through the vibronic coupling $\ket{N_{1,1_0}}\rightarrow \ket{I_{{\tilde{\nu}}}}$, 2) the Ionic species weakly mixes with the QD electronic conduction band $\ket{I_{\tilde{\nu}}}\rightarrow \ket{N_{0,1_k}}$, and 3) the discrete molecular phonon mode interacts with the electronic continuum via the CT mediated by the virtual vibrational ionic intermediate state in a superexchange process $\ket{N_{1,1_0}}\rightarrow \ket{I_{\tilde{\nu}}} \rightarrow \ket{N_{0,1_k}}$. Thus, the Fano asymmetry factor q depends on the electronic coupling between the catalyst and the electronic QD states. To illustrate the superexchange mechanism, Figure \ref{fig:Diagram3} presents a schematic representation. Here, we show the interaction between the discrete vibrational state $\ket{N_{1,1_0}}$ and the electronic continuum of the QD. 

To facilitate the understanding of the fitting parameters for the Fano asymmetry factor $q$ in Eq. \eqref{eq:aprox_q}, we will label the following parameters as $\bar{\mu}_{\t{Ionic}}(t)\equiv \eta_{\nu}^It^2\bar{\mu}_{\nu}^I$ and $W_{\t{QD}}(t) \equiv \bar{W}_{e,C}t^2$, such that is clear in Eq. \eqref{eq:aprox_q} the contributions of the Ionic species, and the QD electronic states:
\begin{align} \label{eq:q_append}
    q & \approx \frac{1}{\pi \rho(\omega_0)} \frac{\bar{\mu}_{10}^N  + \bar{\mu}_{\t{Ionic}}(t)}{W_{\t{QD}}(t) + W_M}
\end{align}
To minimize the number of parameters in our model \eqref{eq:q_append}, we normalize the Fano asymmetry factor by the QD with maximum $q$, i.e., by its maximum value at $t_{\text{max}}$
\begin{align} \label{eq:norm_q}
   \bar{q}(t) &= \frac{q(t)}{q(t_{\text{max}})}= \frac{\left( \frac{W_{\t{QD}}(t_{\t{max}})}{W_M}+1\right)\left( 1+\frac{\bar{\mu}_{\t{Ionic}}(t)}{\bar{\mu}_{10}^N} \right)}{\left( 1 + \frac{\bar{\mu}_{\t{Ionic}}(t_{\t{max}})}{\bar{\mu}_{10}^N} \right) \left( \frac{W_{\t{QD}}(t)}{W_M} + 1 \right)}.
\end{align}
Here, $\frac{\bar{\mu}_{\t{Ionic}}(t_{\t{max}})}{\bar{\mu}_{10}^N}$ is associated with the ratio of the vibrational transition matrix element in the adsorbate ionic state relative to the same quantity for the neutral species of the QD-adsorbate system with electron hopping amplitude $t_{\t{max}}$ and $\frac{W_{\t{QD}}(t_{\t{max}})}{W_M}$ representing the ratio of transition dipole moments for the 1S $\rightarrow$ 1P transition to the vibrational transition matrix element at the resonance frequency of the latter.

\bibliography{References.bib}

\end{document}